\documentclass[longbibliography,aps,reprint,groupedaddress,superscriptaddress]{revtex4-1}

\usepackage[pdftex]{graphicx}
\usepackage{dcolumn}
\usepackage{bm}
\usepackage{amsmath}
\usepackage{latexsym}
\usepackage{epstopdf}
\usepackage{xcolor}
\usepackage{lipsum}
\usepackage{xargs}

\usepackage{lipsum}

\usepackage{hyperref}
\definecolor{bluemy}{rgb}{0.157, 0.173, 0.569}
\definecolor{brickred}{rgb}{0.79, 0.25, 0.33} 
\hypersetup{colorlinks=true, linkcolor=bluemy, citecolor=bluemy, urlcolor=bluemy,anchorcolor=bluemy}

\newcommand{\rfig}[1]{Fig.~\ref{#1}}

\newcommand{\req}[1]{Eq.~(\ref{#1})}

\newcommand{\Rmnum}[1]{\expandafter\@slowromancap\romannumeral #1@}
\newcommand{\XX}{X\!X}

\newcommand{\ZZ}{Z\!Z}

\makeindex

\begin{document}
\title{High-contrast interaction between remote superconducting qubits mediated by multimode cable coupling}
\author{Jiajian Zhang}
\affiliation{Shenzhen Institute for Quantum Science and Engineering, Southern University of Science and Technology, Shenzhen 518055, China}
\affiliation{International Quantum Academy, Shenzhen 518048, China}

\author{Ji Chu}
\email{jichu@iqasz.cn}
\affiliation{International Quantum Academy, Shenzhen 518048, China}

\author{Jingjing Niu}
\email{niujj@iqasz.cn}
\affiliation{International Quantum Academy, Shenzhen 518048, China}
\affiliation{Shenzhen Branch, Hefei National Laboratory, Shenzhen 518048, China}

\author{Youpeng Zhong}
\affiliation{International Quantum Academy, Shenzhen 518048, China}
\affiliation{Shenzhen Branch, Hefei National Laboratory, Shenzhen 518048, China}

\author{Dapeng Yu}
\affiliation{International Quantum Academy, Shenzhen 518048, China}
\affiliation{Shenzhen Branch, Hefei National Laboratory, Shenzhen 518048, China}
\affiliation{School of Physics, Peking University, Beijing 100871, China}

\begin{abstract}
Superconducting quantum processors offer a promising path towards practical quantum computing. However, building a fault-tolerant quantum computer with millions of superconducting qubits is hindered by wiring density, packaging constraints and fabrication yield.
Interconnecting medium‐scale processors via low‐loss superconducting links provides a promising alternative. Yet, achieving high-fidelity two‐qubit gates across such channels remains difficult. 
Here, we show that a multimode coaxial cable can mediate high‐contrast interaction between spatially separated superconducting qubits. 
Leveraging interference between cable modes, we can implement high-fidelity controlled-Z and $\ZZ$-free iSWAP gates by simply modulating qubit frequencies. Numerical simulations under realistic coherence and coupling parameters predict fidelities above 99\% for both gate schemes. 
Our approach provides a versatile building block for modular superconducting architectures and facilitates distributed quantum error correction and large-scale fault-tolerant quantum computing.
\end{abstract}
\maketitle


Quantum error correction (QEC) is a necessary path towards fault-tolerant quantum computation~\cite{nielsen2010quantum,shor1994algorithms,gottesman2002introduction,georgescu2014quantum,gambetta2017building}. Among various quantum computing platforms, superconducting circuits have emerged as one of the leading candidates, with recent demonstration of exponential surface code logical error suppression using approximately 100 physical qubits~\cite{Acharya2024googled7}.
However, practical fault-tolerant quantum computation requires scaling quantum processors to millions of physical qubits~\cite{fowler2012surface,babbush2018encoding,lee2021even}.
Integrating such a large number of qubits within a single superconducting processor remains a substantial challenge due to packaging constraints, wiring complexity, and fabrication yields~\cite{huang2021microwave,Mohseni2024JMsupercomputer}.
Moreover, implementing fault-tolerant quantum algorithms typically requires higher connectivity than nearest-neighbor coupling in planar geometries~\cite{Bravyi2024IBMlowoverhead,Vasmer20193Dsurface}.
Distributed quantum computing, where multiple small-scale quantum processors are connected via high-quality quantum channels, has therefore emerged as a viable path towards large-scale fault-tolerant quantum computers~\cite{jiang2007distributed,gold2021entanglement,ang2024arquin,Leung2019BidirectRemoteEntang}.

Superconducting coaxial cables with high coherence are a key element for connecting spatially separated quantum processors~\cite{Niu2023}, enabling high-fidelity quantum state transfer (QST) between remote qubits~\cite{Zhong2021nat,Axline2018QSTnp,Burkhart2021errordetectQST,Kurpiers2018QSTnat}.
However, the QST process itself does not represent a unitary two-qubit gate, which is typically necessary for distributed QEC. Although a CNOT gate can be synthesized using quantum gate teleportation, such an operation requires complicated feedback control and typically results in low fidelity~\cite{Qiu2025tele}. 
Similar to the bus-mediated coupling architecture, which has been widely used in the field~\cite{Majer2007bus,sillanpaa2007coherent,filipp2011multimode,lucero2012computing,McKay2016tunablebus,takita2017experimental,song201710,cai2018construction,song2019generation,McKay2015CCres,kandala2021demonstration,Zhou2023Pfaffrouter,Heya2025IBMgate}, coaxial cables can be used to directly couple two remote qubits, enabling the implementation of two-qubit gates.
However, fixed coupling generally results in low-contrast qubit interactions, imposing a trade-off between gate speed and infidelity related to parasitic interactions~\cite{barends2014superconducting,krinner2020benchmarking,Zhao2020Cshut}. 
Tunable couplers are widely used in nearest-neighbor coupling to enable high-contrast interaction~\cite{chen2014qubit,yan2018tunable,Goto2022doublecoupler}, thereby enabling high-fidelity two-qubit gates~\cite{foxen2020demonstrating,Xu2020CZ,sung2021realization,Stehlik2021IBM}. These couplers have also been incorporated into bus-mediated or cable-mediated systems to improve the interaction-on/off ratio~\cite{li2019tunable,Zhong2021nat,Zhao2022wide,marxer2023long,Wu2024router,Norris2025performance,Renger2025alltoall}. However, introducing a tunable coupler between the qubit and the cable further weakens the cable-mediated coupling between remote qubits, which hinders the implementation of direct two-qubit gates. Therefore, a high-contrast interaction scheme for cable-mediated coupling without adding extra components is highly desired.

In this work, we analyze the $\ZZ$ and $\XX$ interactions between two remote transmon qubits~\cite{koch2007charge} coupled via a multimode cable. 
We show that $\ZZ$ freedom can be achieved in the near-mode regime, where the qubit-cable detuning is close to the qubit anharmonicity, due to level repulsion from the second excited states of the cable modes. Meanwhile, the $\XX$ interactions mediated by different modes add constructively. 
Leveraging the high-contrast interaction, direct iSWAP and CZ gates between remote qubits can be implemented by modulating qubit frequencies with simple square pulses. Numerical simulations indicate that gate fidelities above 99\% can be achieved by optimizing qubit idle and interaction frequencies under realistic coherence parameters, making distributed QEC feasible. Combined with recent theoretical~\cite{ohfuchi2024remote} and experimental~\cite{Song2024CR,Mollenhauer2024plugplay} studies on cable-mediated Cross-resonance (CR) gates, our work suggests that cable-based remote gates are a reliable approach to distributed QEC, paving the way towards large-scale fault-tolerant quantum computing.

\begin{figure}[t]
    \begin{center}
        \includegraphics[width=0.45\textwidth]{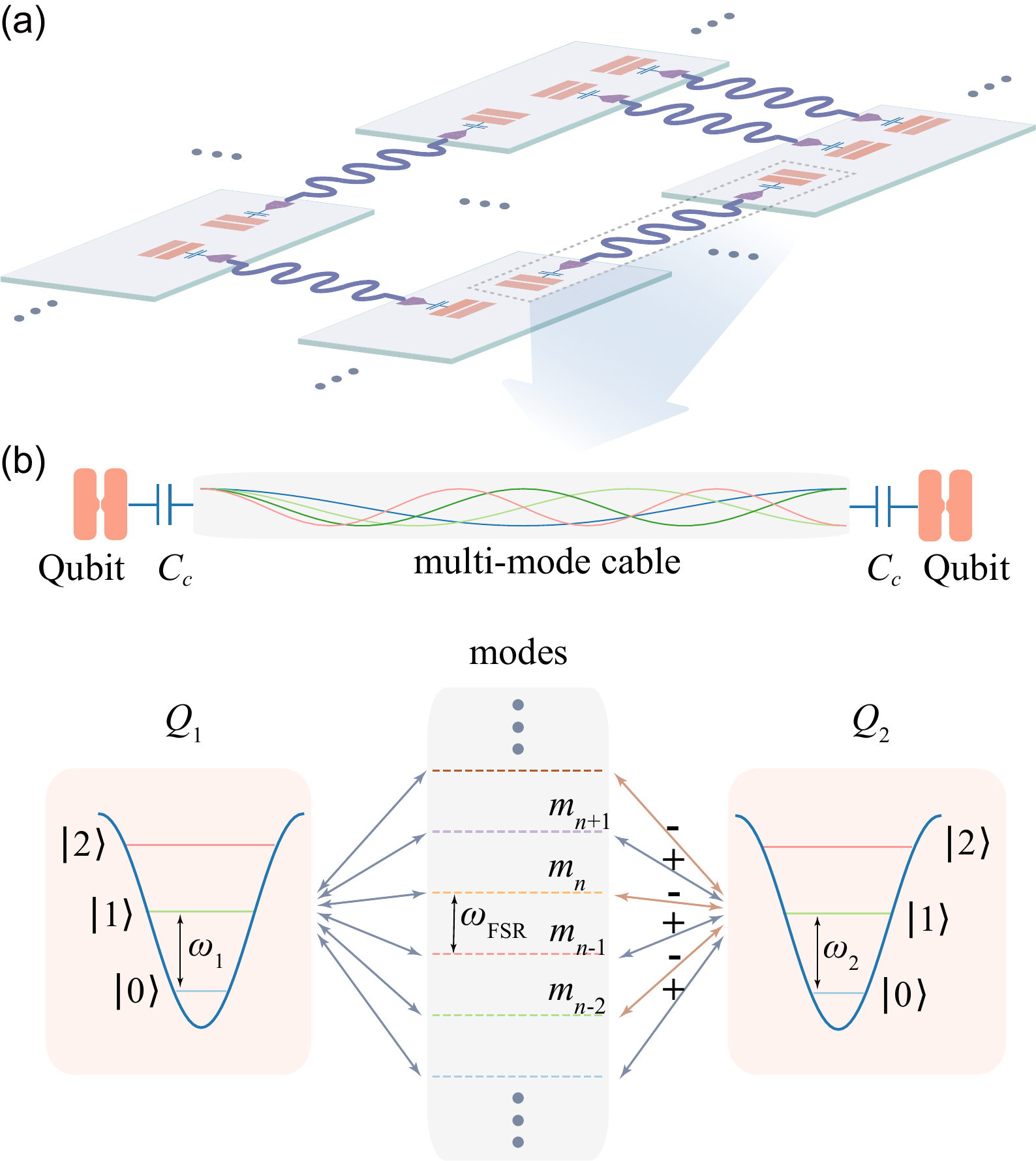}
        \caption{
        \label{Fig1}
        (a) Schematic of distributed superconducting quantum processors. Remote transmon qubits (orange pads) on separate chips are coupled via superconducting coaxial cables (twist lines).
        (b)
        Top panel: Schematic of cable-mediated remote coupling. Multiple standing-wave modes mediate the coupling, with qubits capacitively coupled to the cable between them.
        Bottom panel: Energy-level diagram of the qubits and the cable (middle). 
        Double-headed arrows indicate the coupling between the qubits and cable modes. Due to voltage distribution along the cable, the sign of the coupling strength differs for odd modes (brown: positive coupling; grey: negative coupling).
        \vspace{-10pt}
        }
    \end{center}
\end{figure}

\begin{figure*}[hbt]
    \begin{center}
        \includegraphics[width=0.9\textwidth]{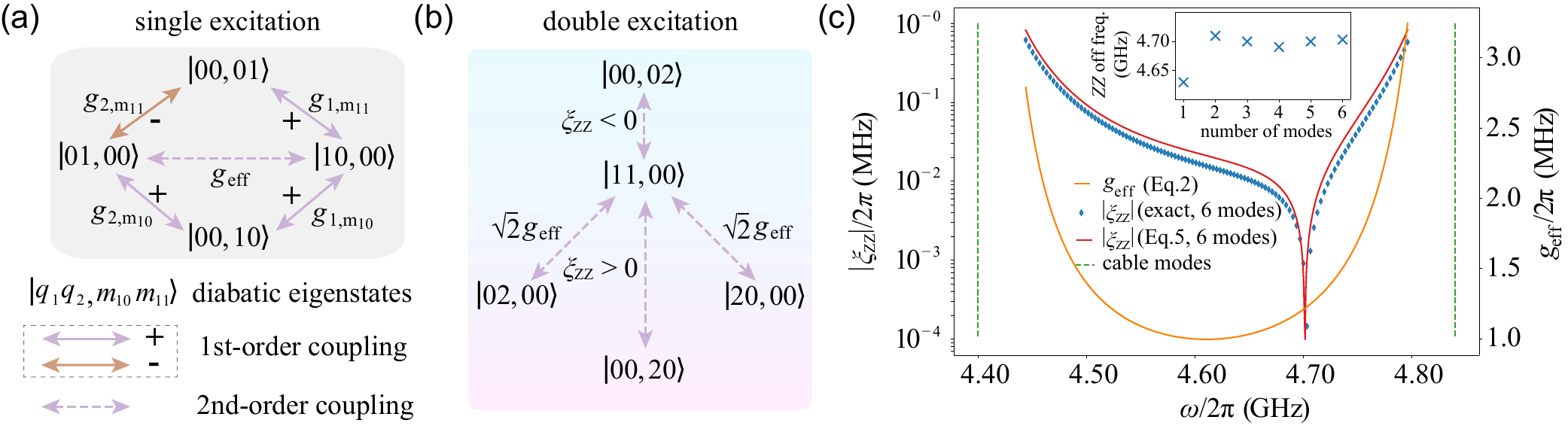}
        \caption{
        \label{Fig2}
        Energy-level diagrams of the single-excitation (a) and double-excitation (b) manifolds. The qubit frequencies are sandwiched between cable modes $m=10$ and $m=11$, with diabatic states labeled $|q_1 q_2, m_{10} m_{11}\rangle$. 
        The symbols $+$ and $-$ indicate the signs of the effective couplings. For the odd mode $m=11$, the coupling to the two qubits have opposite signs due to the spatial voltage distribution along the waveguide.
        In the double excitation manifold, the $\ZZ$ interaction originates from level repulsion between the two-qubit state $|11,00\rangle$ and the second excited states of the qubits and cable modes. 
        (c) The $\XX$ (right axis, orange line) and $\ZZ$ (left axis) interaction strengths as functions of qubit frequencies between modes $m=10$ and $m=11$, with a FSR of $\omega_{\rm FSR} = 440$ MHz. The qubit frequencies are assumed to be near resonance. 
        The exact diagonalization result for the $\ZZ$ interaction strength (blue diamonds) is compared to the one calculated from the analytical formula in \req{eqn:zzoffpoint_approx} (red line).
        The green dashed lines indicate frequencies of the cable modes.
        The inset shows the $\ZZ$ frequency as a function of the number of modes included in the diagonalization calculation.
        \vspace{10pt}
                }
    \end{center}
\end{figure*}

\vspace{5pt}
\textbf{Cable-mediated coupling}
\vspace{5pt}

We consider a distributed quantum processor (\rfig{Fig1}{(a)}) comprising multiple superconducting chips connected by superconducting coaxial cables, as illustrated in \rfig{Fig1}(a).
The open ends of each cable are capacitively coupled to two qubits located on separate chips. 
We focus on the minimal configuration where two qubits $Q_1$ and $Q_2$ are coupled to a multimode cable, as shown in \rfig{Fig1}(b). The cable acts as a linear oscillator with discrete modes separated by the free spectral range (FSR) $\omega_{\mathrm{FSR}}$. 
The system Hamiltonian (in units of $\hbar$ hereafter) is given by
\begin{equation}\small\label{Hami}
\begin{split}
    H\! &=
\!\sum_{i=1,2}\!\big[\omega_{i}a_{i}^{\dagger}a_{i}\!+\!\frac{1}{2}\alpha_{i}a_{i}^{\dagger}a_{i}^{\dagger}a_{i}a_{i}\! \big]\! +\!\sum_{m}\!m\omega_{\mathrm{FSR}}c_{m}^{\dagger}c_{m}\! \\ 
& + \!\sum_{m}\big[g_{1,m}(a_{1}^{\dagger}c_{m}\!+\!c_{m}^{\dagger}a_{1}) + \!(-1)^{m}g_{2,m}(a_{2}^{\dagger}c_{m}\!+\!c_{m}^{\dagger}a_{2})\big],
\end{split}
\end{equation}
\noindent
where $a_i$ ($c_m$) annihilates an excitation in $i$-th qubit ($m$-th cable mode), $\omega_i$ and $\alpha_i$ are the qubit frequency and anharmonicity, $\omega_{\rm FSR}$ is the cable’s FSR, and $g_{i,m}$ is the transverse coupling strength of the qubit to cable mode $m$. Importantly, due to the spatial voltage distribution along the half-wavelength waveguide, the odd- and even-indexed modes couple to $Q_2$ with opposite signs (assuming positive coupling to $Q_1$).

In this work, we assume a cable length of 0.25~m, corresponding to an FSR of $\omega_{\rm FSR}/2\pi = 440$~MHz. For simplicity, we analyze the $\ZZ$ interaction strength when the two qubit frequencies are near resonance and located between two selected modes $m=10$ and $m=11$.
\rfig{Fig2}(a) and (b) sketch the energy level diagrams of single- and double-excitation manifolds, showing the diabatic states in the form of $|q_1 q_2, m_{10} m_{11}\rangle$. The diabatic states are obtained by neglecting all the coupling terms in \req{Hami}~\cite{chu2021coupler,Li2024doublecoupler}. 
The effective $\XX$ interaction between two qubits mainly arises from second-order couplings mediated by the two nearest cable modes. 
The two interaction paths add constructively due to the opposite (same) signs of the coupling strengths to the two qubits mediated by the odd (even) modes.
Using the Schrieffer-Wolff transformation, we obtain the $\XX$ interaction strength $g_{\rm eff}$ as:
\begin{equation}\small\label{XX_interaction}
    \begin{split}
    g_{\rm{eff}} \approx \frac{1}{2}\sum_{m=1}^{\infty} (-1)^m g_{1,m}g_{2,m} \left(\frac{1}{\Delta_{1,m}}+\frac{1}{\Delta_{2,m}}  
    - \frac{1}{\Sigma_{1,m}}-\frac{1}{\Sigma_{2,m}}\right)
    \end{split}
\end{equation}
where $g_{i,m}$ denotes the coupling between qubit $i$ and the $m$-th cable mode, given by $g_{i,m}=\frac{1}{2}\frac{C_{\rm{c}}}{\sqrt{C_i C_m}}\sqrt{\omega_i \omega_m}$, with $C_{\mathrm{c}}$ the coupling capacitance, and $C_i$, $C_m$ the qubit and cable self capacitances, respectively. The detunings are defined as $\Delta_{i,m}=\omega_i-m\omega_{\mathrm{FSR}}$ and $\Sigma_{i,m}=\omega_i+m\omega_{\mathrm{FSR}} \ (i=1,2)$. 
In the numerical simulation presented in the main text, the following parameters are used: $C_q = 90$~fF, corresponding to an anharmonicity of $-216$~MHz; $C_c = 5$~fF, and the self capacitance of the 0.25~m cable is $C_m = 11.75$~pF.
For variations in these parameters, refer to \cite{SM2025} for details.

\vspace{20pt}
\textbf{High contrast $\ZZ$ interaciton}
\vspace{10pt}

The cable modes facilitate a high on-off ratio of the $\ZZ$ interaction strength between the two qubits. The $\ZZ$ interaction, denoted by $\xi_{\rm{ZZ}}$, is defined as
\begin{equation}
\xi_{\rm{ZZ}} = \tilde{\omega}_{11,00} + \tilde{\omega}_{00,00}-\tilde{\omega}_{10,00}-\tilde{\omega}_{01,00},
\end{equation}
where $\tilde{\omega}$ represents the frequency of the corresponding eigenstate in the qubit-cable-qubit system.
Using fourth-order perturbation theory, we derive an analytical expression for the $\ZZ$ interaction strength:
\begin{widetext}
\begin{equation}
\label{eqn:xi_zz_complete}
\begin{split}
\xi_{\mathrm{ZZ}} &\approx \sum_{m=1}^{\infty} g_{1,m}^2 g_{2,m}^2 \Bigg\{ \frac{1}{\Delta_{12}} \left[ \left( \frac{1}{\Delta_{2,m}} - \frac{1}{\Sigma_{1,m}} \right)^2 - \left( \frac{1}{\Delta_{1,m}} - \frac{1}{\Sigma_{2,m}} \right)^2 \right] \\
&\quad + \frac{2}{\Delta_{12} - \alpha_2} \left( \frac{1}{\Delta_{1,m}} - \frac{1}{\Sigma_{2,m} + \alpha_2} \right)^2 - \frac{2}{\Delta_{12} + \alpha_1} \left( \frac{1}{\Delta_{2,m}} - \frac{1}{\Sigma_{1,m} + \alpha_1} \right)^2 \\
&\quad + \frac{1}{\Sigma_{12} - 2 m \omega_{\mathrm{FSR}}} \left[ 2 \left( \frac{1}{\Delta_{1,m}} + \frac{1}{\Delta_{2,m}} \right)^2 - \frac{1}{\Delta_{1,m} \Delta_{2,m}} \right] \Bigg\},
\end{split}
\end{equation}
\end{widetext}
\noindent
where $\Delta_{12}=\omega_1-\omega_2$ and $\Sigma_{12}=\omega_1+\omega_2$.

The $\ZZ$ interaction strength arises primarily from nontrival level repulsions involving the second-excited states of the qubits and cable modes, as illustrated in \rfig{Fig2}(b) (see also \cite{SM2025}). 
To analyze the conditions for achieving $\ZZ$ freedom, we focus on the near-resonance regime where $\Delta_{12} \to 0$ and assume $\alpha_1=\alpha_2 = \alpha$. 
Under these conditions, the approximate form of $\xi_{\mathrm{ZZ}}$ simplifies to
\begin{equation}\label{eqn:zzoffpoint_approx}
    \begin{split}
    \xi_{\rm{ZZ}} &\approx \sum_{m=1}^{\infty} {g_{1,m}^2 g_{2,m}^2}\times\\
    &\left[-\frac{4}{\Delta_{m}^2\alpha} + \frac{11}{2\Delta_m^3} + \frac{8}{\Delta_m\alpha(\Sigma_m+\alpha)} -\frac{2}{\Delta_m^2 \Sigma_m} \right], 
    \end{split}
\end{equation}
where $\Delta_{m}=\omega-m\omega_{\rm{FSR}}$ and $\Sigma_{m}=\omega + m\omega_{\rm{FSR}}$. The first and second terms in \req{eqn:zzoffpoint_approx} arise from level repulsions associated with the second-excited states of the qubits and the cable modes, respectively. 
$\ZZ$-free operation can be achieved at specific qubit frequencies where the negative level repulsions from higher-frequency cable modes cancels out the positive contributions from other energy levels. 
\rfig{Fig2}(c) displays the calculated $\XX$ and $\ZZ$ interaction strengths as functions of the average qubit frequency, assuming the two qubits are near resonance. The two nearest cable modes contribute most significantly to the level repulsions, and the calculated $\ZZ$ freedom frequency converges when two cable modes are considered, as shown in the inset of \rfig{Fig2}(c).
High-contrast qubit interactions are realized at the $\ZZ$-free point because the $\XX$ interactions mediated by adjacent cable modes add constructively. Such high-contrast interactions enable strong qubit-cable coupling, thereby allowing direct interaction between two remote qubits.

\begin{figure*}[t]
    \begin{center}
        \includegraphics[width=0.9\textwidth]{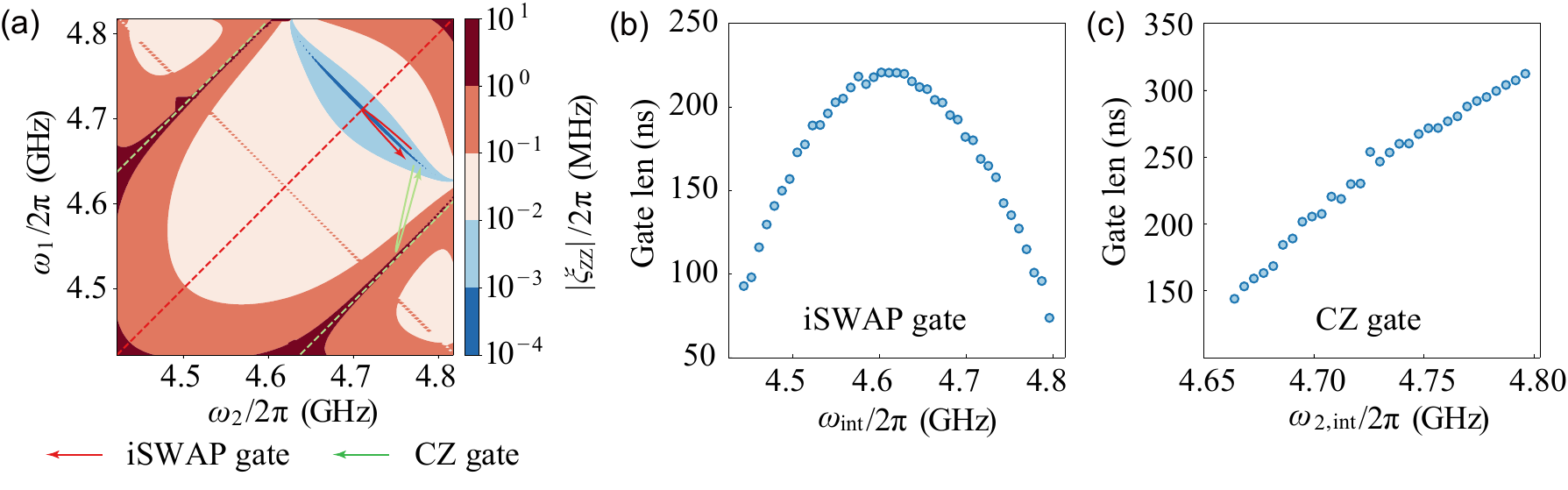}
        \caption{
            \label{Fig3gatescheme}
            {(a) Absolute $\ZZ$ interaction strength $|\xi_\mathrm{ZZ}|$ versus $Q_1$ and $Q_2$ frequencies. The blue region indicates the qubit idling area where $|\xi_\mathrm{ZZ}|/2\pi<10$~kHz. 
            The red (green) dashed line marks interaction frequencies for iSWAP (CZ) gates. The red (green) curve with an arrow illustrates the frequency trajectory during the iSWAP (CZ) gate. Square pulses are used for qubit frequency modulations. (b) Simulated iSWAP gate length versus interaction frequency $\omega_\mathrm{int}$. (c) Simulated CZ gate length versus $Q_2$ interaction frequency ($\omega_\mathrm{2,int}$), with interaction frequency of $Q_1$ detuned by an anharmonicity. 
            }
        }
    \end{center}
\end{figure*}

\vspace{10pt}
\textbf{Remote iSWAP and CZ gates}

In the analysis of the $\ZZ$ interaction strength presented above, we assume the qubit frequencies to be near resonance for simplicity. However, in actual gate implementations, the qubit frequencies must be sufficiently detuned ($\Delta_{12} \gg g_{\rm eff} $) to minimize state hybridization during idle periods. \rfig{Fig3gatescheme}(a) presents the full $\ZZ$ interaction spectrum as a function of the two-qubit frequencies, spanning the region between cable modes $m=10$ and $m=11$.
We choose the idle frequencies in the blue region, where the absolute $\ZZ$ interaction strength remains below 10~kHz, thereby suppressing unwanted interactions.
To implement two-qubit gates, we dynamically tune the system into resonance.
The iSWAP gate is implemented by tuning the states $|10,00\rangle$ and $|01,00\rangle$ into resonance, as indicated by the red dashed line. 
Similarly, the CZ gate is realized by tuning the $|11,00\rangle$ and $|20,00\rangle$ (or $|02,00\rangle$) states into resonance, as shown by the green dashed line.
For simplicity, we focus here on gate performance under square-pulse frequency modulations for both qubits (results for alternative waveforms are provided in \cite{SM2025}).
\rfig{Fig3gatescheme}(b, c) presents the iSWAP and CZ gate durations, respectively, as functions of the interaction frequencies. For the iSWAP gate, the gate speed increases as $\omega_\mathrm{int}$ approaches a cable mode, due to the stronger $\XX$ interaction near cable modes.
In contrast, the CZ gate achieves shorter durations when the interaction frequency of $Q_2$ ($\omega_\mathrm{2,int})$ is close to the lower-frequency cable mode (with $Q_1$ detuned from $Q_2$ by the anharmonicity $\alpha$), since the $\ZZ$-off region is close to the higher-frequency mode.  

\begin{figure*}[!hbtp]
    \begin{center}
        \includegraphics[width=0.9\textwidth]{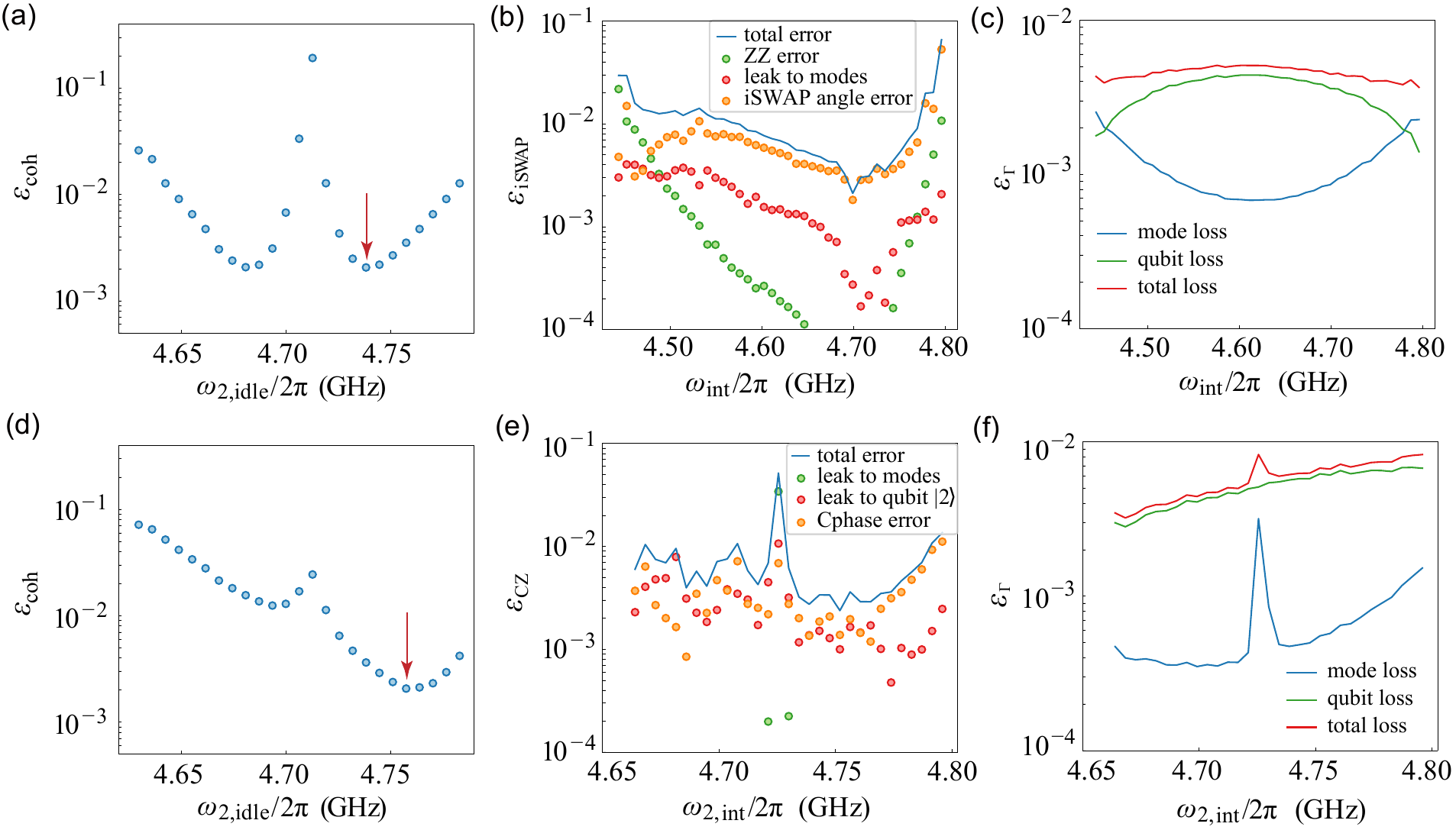}
        \caption{
            \label{CZ_error}
            {Error analysis of remote iSWAP and CZ gates.}
            (a) Coherent error of iSWAP gates as a function of the idling frequency of $Q_2$. The idling frequency of $Q_1$ varies accordingly, extracted from the $\ZZ$-off region in \rfig{Fig3gatescheme}{(a)}. The interaction frequency of the two qubits is 4.708~GHz, corresponding to the resonant $ZZ$ freedom point.
            The red arrow indicates the optimal idling frequency of $Q_2$. 
            (b) Coherent errors of optimized iSWAP gates as functions of the interaction frequency, including: leakage errors (red circles), conditional phase errors (green circles), and iSWAP angle errors (orange circles). The blue line indicates the total coherent error.
            (c) Incoherent errors of the iSWAP gates versus interaction frequency, assuming qubit and cable mode lifetimes of $100~\mu$s and $10~\mu$s, respectively. 
            (d) Coherence error of CZ gates versus the idling frequency of $Q_2$, with the idling frequency of $Q_1$ extracted from the $\ZZ$-off region in \rfig{Fig3gatescheme}{(a)}. The error peak at 4.72~GHz corresponds to the two-qubit resonance. 
            The interaction frequencies are assumed to be 4.75~GHz and 4.54~GHz for $Q_1$ and $Q_2$, respectively.
            (e) Coherent errors versus interaction frequency of $Q_2$, with the interaction frequency of $Q_1$ detuned by an anharmonicity. The error peak corresponds to the second-order coupling between $|11,00\rangle-|00,11\rangle$.
            (f) Incoherent errors of CZ gates as a function of interaction frequency of $Q_2$, assuming qubit and cable mode lifetimes of $100~\mu$s and $10~\mu$s, respectively. 
            }
    \end{center}
\end{figure*}

For the diabatic frequency control used in our gate scheme, the idle frequencies significantly impact the gate fidelities due to state hybridization. In \rfig{CZ_error}{(a)}, we analyze the coherent gate error as a function of the idle frequency of $Q_2$, with the interaction frequency of both qubits set to 4.708~GHz. The idle frequency of $Q_1$ is selected to minimize the $\ZZ$ interaction, as determined from \rfig{Fig3gatescheme}{(a)}. 
For each data point, the gate parameters are optimized by maximizing the unitary fidelity~\cite{Kueng2016}, defined as:
\begin{equation}
    \mathcal{F} \equiv (|tr(U_{\rm{ideal}}^{\dagger} U_{\rm{simu}})|^2 + tr(U_{\rm{simu}}^{\dagger} U_{\rm{simu}}))/20.\nonumber
    \label{eqn:fid}
\end{equation}
The corresponding coherent gate error is then given by $\varepsilon_{\rm coh} = 1 - \mathcal{F}$.
Higher gate errors are observed when one of the idle frequencies approaches a cable mode, or when the two qubits are near resonance. This indicates that state hybridization--whether between the qubits themselves or between the qubits and the cable modes--fundamentally limits the control fidelity. The optimized idle frequency for $Q_2$ is 4.738~GHz, with the corresponding $Q_1$ idle frequency at 4.684~GHz.
The coherent errors of iSWAP gates mainly arise from three parts: leakage error to the cable modes, iSWAP angle error, and conditional phase error accumulated during the interaction~\cite{Eickbusch2024dynamicCode}, as shown in \rfig{CZ_error}(b).
We find that the minimal $\ZZ$ error point coincides with the minimal leakage error point, where the total control error is optimized to 0.19\%.
For incoherent errors, we consider the relaxation processes associated with qubit and cable loss. The relaxation times for the bare qubits and cable modes are assumed to be 100~$\mu$s and 10 $\mu$s, respectively. We estimate the loss of the involved eigenstates in the logical subspace by intergrating the overlap with bare qubit and cable modes during the gate operation.
The incoherent errors $\varepsilon_{\rm{\Gamma}}$ versus the interaction frequencies are shown in \rfig{CZ_error}{(c)}. At the optimized interaction frequency of $\omega_{\rm{int}}/2\pi=4.708$ GHz, the total error of the 180~ns iSWAP gate is estimated to be $0.62\%$, comprising 0.31\% qubit loss, 0.12\% cable mode loss, and 0.19\% coherent errors.

For CZ gates, the qubit frequencies are tuned from the $\ZZ$ off region into $|11,00\rangle-|20,00\rangle$ or $|11,00\rangle-|02,00\rangle$ resonance, as indicated by the green dashed lines in \rfig{Fig3gatescheme}{(a)}.
Similar to the iSWAP gate, we optimize the idle frequencies of the qubits in \rfig{CZ_error}(d) by fixing the interaction frequencies at $\omega_{\rm{q1,int}}/2\pi=4.54$ GHz and $\omega_{\rm{q2,int}}/2\pi=4.75$ GHz. 
With the optimal idling frequencies $\omega_{\rm{q1,idle}}/2\pi=4.665$ GHz and $\omega_{\rm{q2,idle}}/2\pi=4.758$ GHz, the interaction frequency is then optimized in \rfig{CZ_error}(e). 
We show that the coherent error of a diabatic CZ gate is mainly attributed to leakage to second excited states of the qubits and conditional phase errors, due to imperfect $|11,00\rangle-|02,00\rangle$ state swaps.  
The total coherent error based on diabatic control can be suppressed to 0.2\% with optimized idle and interaction frequencies. 
Note that the coherent error can be further reduced to 0.03\% using an optimized adiabatic control pulse~\cite{Martinis2014fastadia}, but this results in a longer gate time~\cite{SM2025}. 
At the optimal interaction frequency, the incoherent error is estimated to be $\varepsilon_{\rm{\Gamma}}=0.68\%$, assuming qubit $T_1$ of 100~$\mu$s and cable mode $T_1$ of $10~\mu$s.
The extreme values in \rfig{CZ_error}{(e)} and \rfig{CZ_error}{(f)} originate from the resonance of $|11,00\rangle$ and $|00,11\rangle$, leading to significant state leakage to the cable modes.
Taking all coherent and incoherent errors into account, a remote diabatic CZ gate with a gate time of 271~ns can achieve a fidelity of $99.14\%$.

\vspace{10pt}
\textbf{Discussion and Outlook}

In conclusion, we investigate high-contrast interaction between remote superconducting qubits facilitated by a multimode coaxial cable. 
The high-contrast interaction arises from constructive interference of $\XX$ interaction and destructive interference of the $\ZZ$ interaction, with $\ZZ$ freedom achievable in the near-mode regime.
By optimizing the qubit idle and interaction frequencies, high fidelity iSWAP and CZ gates can be implemented using simple square pulses under realistic coherence parameters.
The coherent control error can be further suppressed by optimizing the control pulses, at the cost of increased gate time~\cite{SM2025}.
Together with recent advancements in cable-mediated Cross-resonance (CR) gates~\cite{Song2024CR,Mollenhauer2024plugplay}, our work underscores that cable-based distributed quantum processor is a vital approach towards large-scale fault-tolerant quantum computing.

\begin{acknowledgments}
\vspace{-10pt}
This work was supported by the National Natural Science Foundation of China (12374474, 12174178, 123b2071), the Science, Technology and Innovation Commission of Shenzhen Municipality (KQTD20210811090049034, K21547502), the Innovation Program for Quantum Science and Technology (2021ZD0301703), and the Guangdong Basic and Applied Basic Research Foundation (2022A1515110615, 2024A1515011714).
\end{acknowledgments}

%
 
%

\end{document}


\title{Supplemental Material for ``High-contrast interaction between remote superconducting qubits mediated by multimode cable coupling''}

\maketitle

\setcounter{equation}{0}
\setcounter{figure}{0}
\setcounter{table}{0}
\setcounter{page}{1}

\tableofcontents
\newpage

\section{Full energy spectrum}

\begin{figure*}[hbt]
    \begin{center}
        \includegraphics[width=0.9\textwidth]{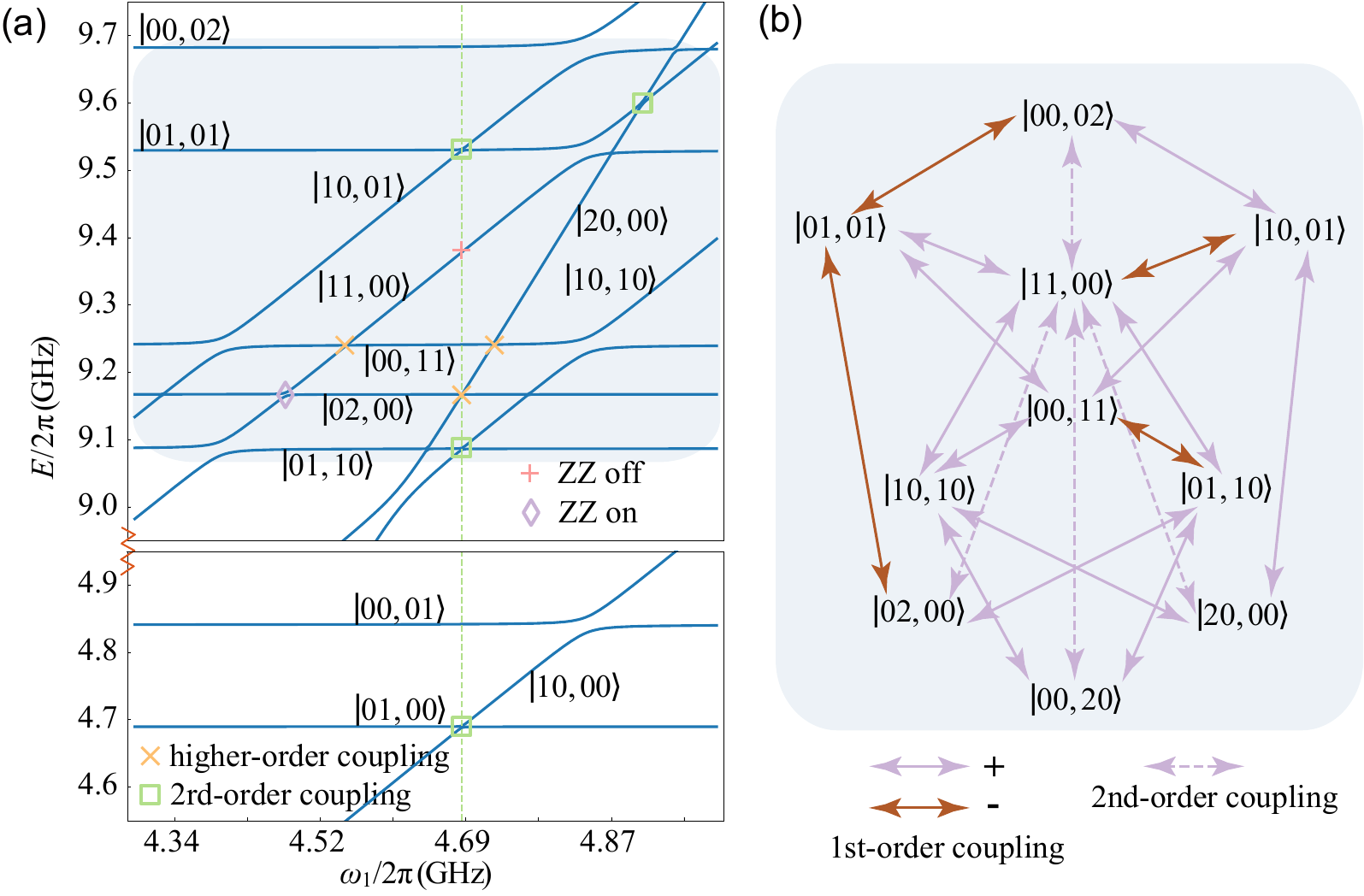}
        \caption{
            \label{fullspec}
            {Energy spectrum of the coupled qubit-cable-qubit system. (a) Energy spectrum versus $Q_1$ frequency $\omega_1$. The frequency of $Q_2$ is fixed at $\omega_2/2\pi=4.70$~GHz. The green dashed line indicates the $ZZ$-off regime. (b) The energy level diagram of the two-excitation manifold, with two qubits tuned near resonance. The dashed lines indicate second-order interactions between $\ket{11,00}$ and the second excited states: $|00,02\rangle$ , $\ket{02,00}$, $\ket{00,20}$ and $\ket{20,00}$. }
        }
    \end{center}
\end{figure*}

Figure~\ref{fullspec}(a) presents the calculated energy spectrum of the qubit-cable-qubit system versus $Q_1$ frequency $\omega_1$, with $Q_2$ frequency fixed at $\omega_2/2\pi=4.7$~GHz. Two cable modes, $m=10$ and $m=11$, are considered. The states are labeled as $\ket{q_1 q_2, m_{10}m_{11}}$ and the shaded region emphasizes the two-excitation manifold.

Strong $\ZZ$ interaction is turned on near the $\ket{11,00}\leftrightarrow\ket{02,00}$ or $\ket{11,00}\leftrightarrow\ket{20,00}$ resonance. 
$\ZZ$ freedom arises from the balance of level repulsions from the second excited states. Note that only level repulsions from the second excited states are non-trivial.
As illustrated in Fig.~\ref{fullspec}(b), $\ZZ$ freedom primarily results from the downward repulsion of the $|00,02\rangle$ state on $|11,00\rangle$, competing with the upward repulsion by $\ket{02,00}$, $\ket{00,20}$ and $\ket{20,00}$. The interactions between the second excited states and the state $\ket{11,00}$ are second-order interactions mediated by states like $\ket{01,01}$.
Consequently, the $ZZ$ freedom point is closer to the higher cable-mode frequency, and the $\ZZ$ coupling strength increases as the qubit frequencies approach the lower cable-mode frequency.

\section{$\ZZ$ interaction}

\begin{figure*}[hbt]
    \begin{center}
        \includegraphics[width=0.8\textwidth]{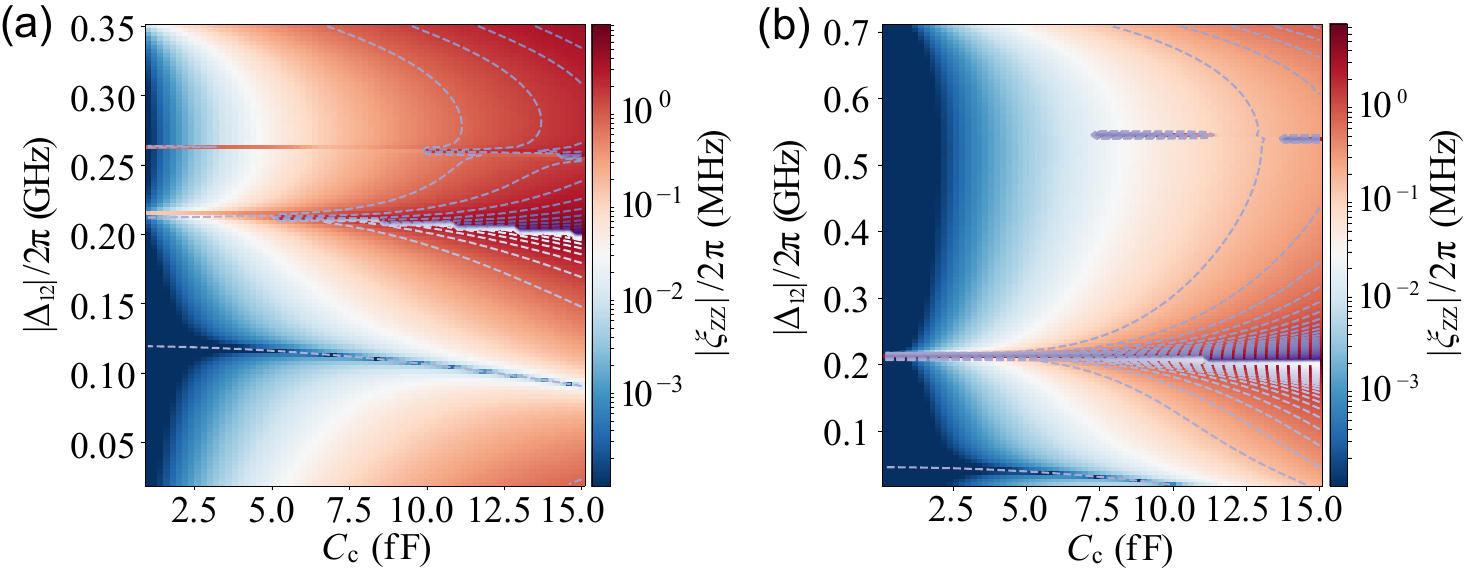}
        \caption{
            \label{ZZCcDelta}
            {Absolute $\ZZ$ interaction strength versus qubit-cable coupling capacitance $C_c$ and qubit-qubit detuning $\Delta_{12}$. (a) $\omega_{\rm{FSR}}/2\pi=440$ MHz (cable length 0.25~m) and (b) $\omega_{\rm{FSR}}/2\pi=917$ MHz (cable length 0.12~m). }
        }
    \end{center}
\end{figure*}
Figure~\ref{ZZCcDelta}{(a)} shows the absolute $\ZZ$ interaction strength $|\xi_{\rm ZZ}|$ versus qubit-cable coupling capacitance $C_c$ and qubit–qubit detuning $\Delta_{12}=\omega_1-\omega_2$ (with $\omega_2/2\pi=4.752$\,GHz). As $C_c$ increases, the $\ZZ$ freedom point moves to smaller $\Delta_{12}$. If $C_c$ becomes too large, the system enters the non-dispersive regime, degrading single-qubit gates and readout fidelities. 
In the main text, we evaluate the gate performance for $C_c=5$ fF, corresponding to a coupling strength of $g_{q,m}/2\pi\sim 11.5$~MHz. 
\rfig{ZZCcDelta}(b) presents a similar scan for a larger free-spectral range, $\omega_{\rm FSR}/2\pi=917$\,MHz (cable length 0.12 m), confirming that the $ZZ$ freedom persists across realistic cable lengths. With a shorter cable length, the actual coupling strength $g_{q,m}$ increases, and the $ZZ$ freedom points move to a smaller $\Delta_{12}$. 

\begin{figure*}[hbt]
    \begin{center}
        \includegraphics[width=0.8\textwidth]{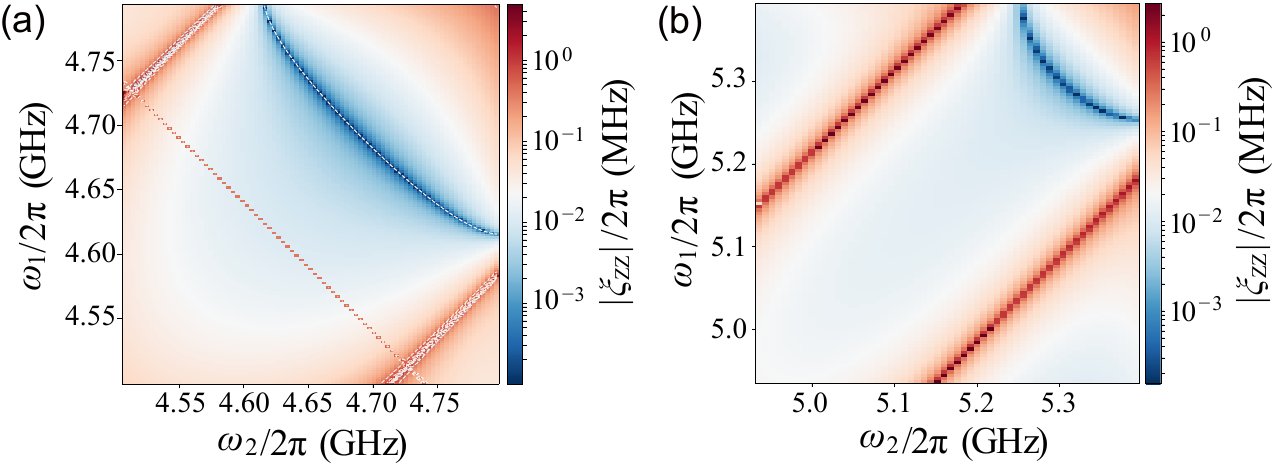}
        \caption{
            \label{ZZ_w1w2}
            {Absolute $\ZZ$ interaction strength versus qubit frequencies $\omega_{1}$ and $\omega_{2}$. (a) $\omega_{\rm{FSR}}/2\pi=440$~MHz and (b) $\omega_{\rm{FSR}}/2\pi=917$~MHz. 
            }
        }
    \end{center}
\end{figure*}

In \rfig{ZZ_w1w2}, we show the absolute $\ZZ$ interaction strength $|\xi_{\rm ZZ}|$ as a function of qubit frequencies $\omega_1$ and $\omega_2$ for $\omega_{\rm FSR}/2\pi=440$\,MHz (\rfig{ZZ_w1w2}{(a)}) and 917\,MHz (\rfig{ZZ_w1w2}{(b)}). At $\omega_{\rm FSR}/2\pi=440$\,MHz, the denser mode spectrum involves multiple-mode mediated coupling, yielding a nearly ``straight'' $\ZZ$-off contour (\rfig{ZZ_w1w2}{(a)}). By contrast, the broader spacing (917\,MHz)  effectively isolates a single dominant mode, resulting in a noticeably curved $\ZZ$ off trajectory (\rfig{ZZ_w1w2}{(b)}). 
Although shorter cables (larger $\omega_{\rm FSR}$) can enhance coupling strength, the 440~MHz configuration is advantageous experimentally. Its straighter $\ZZ$-off path simplifies gate optimization, reducing the need for fine‐tuned iSWAP trajectories to suppress unwanted $\ZZ$ interactions. It also reduces the required flux‐pulse amplitude and confines the qubit tuning range, thereby simplifying pulse calibration and reducing hardware complexity. Consequently, the 440~MHz FSR represents a compromise between coupling strength and practical implementation.

\begin{figure*}[hbt]
    \begin{center}
        \includegraphics[width=0.45\textwidth]{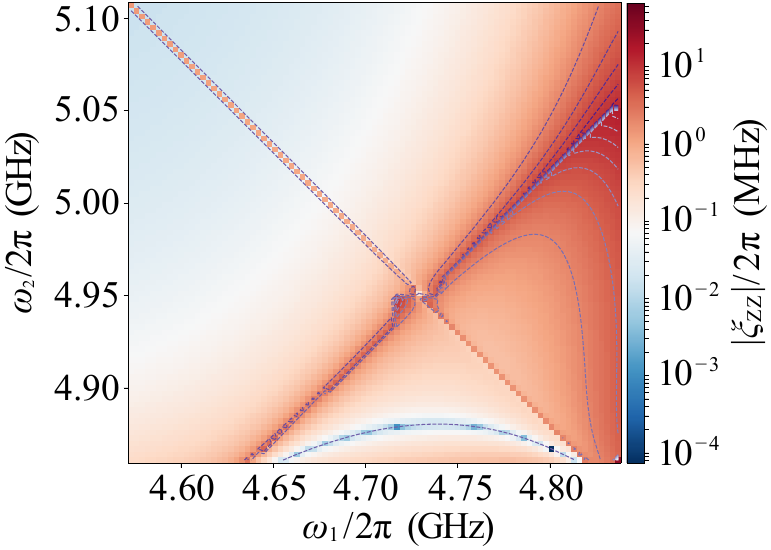}
        \caption{
            \label{ZZ_crossmode}
            {Cross-mode $\ZZ$ interaction strength versus frequencies of two qubits ($\omega_1$ and $\omega_2$), where $\omega_1 < 11\times \omega_{\rm{FSR}}$, $\omega_2 > 11\times\omega_{\rm{FSR}}$, with $\omega_{\rm{FSR}}/2\pi=440$ MHz. }
        }
    \end{center}
\end{figure*}
The high-contrast $\ZZ$ interaction also exists between qubits with frequencies across a cable mode. We calculate the $\ZZ$ spectrum for this case, with $Q_1$ frequency positioned below and $Q_2$ frequency above the 11-th cable mode (see \rfig{ZZ_crossmode}). 
This feature expands the range of feasible designs for parallel gate operations and $\ZZ$-free parametric gates across cable modes~\cite{Mollenhauer2024plugplay}.

\section{Pulse optimization for CZ gates}

\begin{figure*}[hbt]
    \begin{center}
        \includegraphics[width=0.9\textwidth]{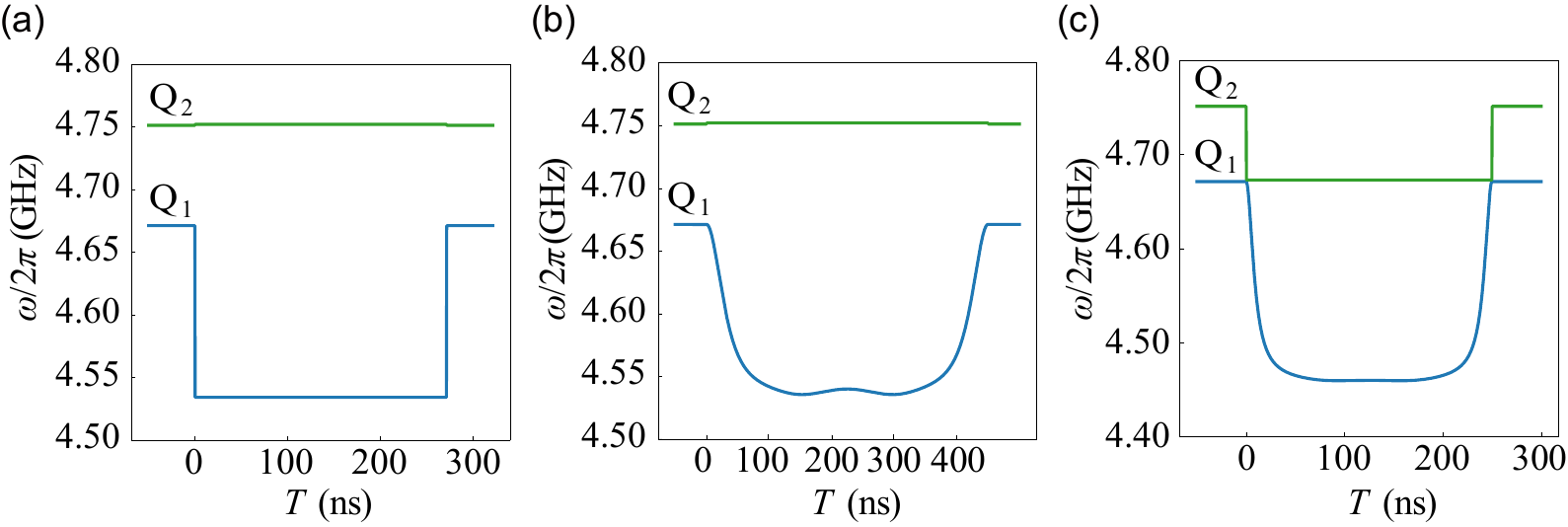}
        \caption{
            \label{pulse_opt}
            {Pulse-shape optimization of CZ gates. (a) Square pulses on both qubits for optimizing interaction frequencies. (b) Optimized Slepian pulses for the same optimal interaction frequencies. (c) Fast CZ gate combining a square pulse on one qubit and an optimized Slepian pulse on the other.}
        }
    \end{center}
\end{figure*}

In the main text, we employ square flux pulses to implement all gates, as they offer speed and scalability when modulating a large number of qubits simultaneously. However, due to the weak anharmonicity of transmon qubits, square-pulse CZ gates inevitably suffer from a trade-off between leakage and conditional phase errors. To address this, we explore Slepian-based pulse shaping~\cite{Martinis2014fastadia,sung2021realization} as a strategy to suppress leakage while maintaining accurate conditional phase angle. Define 
\begin{equation}
\gamma(t) = 0.5(\theta_f-\theta_i)\sum \lambda_k \left(1-\cos(2(k+1)\pi t/\tau) \right),
\label{eqn:slepian}
\end{equation}
\noindent
where $\theta_i$ and $\theta_f$ denote the initial and final angles, respectively, $\lambda_k$ is the k-th order Fourier coefficient of the pulse, $\tau$ represents the gate length. 
The time-domain shape of the optimized Slepian pulse can be expressed as:
\begin{equation}
\Delta\omega(t)=\omega_{\rm{int}}-\omega_{\rm{idle}}+\frac{2 J_{11,02}}{\tan[\gamma(t)]},\nonumber
\end{equation}
where $J_{11,02}$ denotes the coupling strength between the $|11,00\rangle$ and $|02,00\rangle$ states, and $\gamma(t)$ can be expanded to multiple orders as given in \req{eqn:slepian}. 
The relation between the initial angle and the idle/interaction frequencies can be described as: $\tan(\theta_i)=2J_{11,02}/(\omega_{\rm{idle}}-\omega_{\rm{int}})$. 
For practical implementation, we adopt a third-order Slepian expansion, which ensures smooth pulse turn-off at the gate’s end. The optimal parameters for a 450~ns gate are found to be $\lambda_1 \approx 1.273$, $\lambda_2 \approx 0.550$, $\lambda_3 \approx -0.273$, and $\theta_f \approx 0.449\pi$.

At the optimal idle and interaction frequencies, the square-pulse CZ gate (\rfig{pulse_opt}{(a)}) yields a gate duration of 271 ns with a control error of 0.2$\%$. By contrast, using a Slepian-optimized CZ pulse at the same frequencies reduces the control error to 0.03\%, with an increased gate duration of 450~ns (\rfig{pulse_opt}{(b)}). 
If qubit coherence permits, this optimized pulse is preferred.
Alternatively, we note that the total gate duration can be reduced below 250~ns by applying Slepian-based modulation to one qubit while retaining square-pulse modulation on the other (\rfig{pulse_opt}{(c)}). This hybrid approach still achieves a respectable control error of 0.4\%, offering an trade-off between gate speed and fidelity.


%